\documentclass{article}
\usepackage{spconf,amsmath,graphicx}
\usepackage{subcaption}
\usepackage{amssymb}
\usepackage{amsfonts}
\usepackage{mathtools}
\usepackage{multirow}
\usepackage{bm}
\usepackage{nicefrac}
\usepackage{tabularx}
\usepackage{booktabs}
\usepackage{pifont}
\usepackage{array}
\usepackage{balance}

\DeclareMathOperator*{\argmin}{argmin}

\makeatletter
\def\hlinewd#1{%
  \noalign{\ifnum0=`}\fi\hrule \@height #1 \futurelet
   \reserved@a\@xhline}
\makeatother
\title{IMPROVING UNIVERSAL SOUND SEPARATION USING SOUND CLASSIFICATION}
\name{Efthymios Tzinis$^{1,2,*}$\thanks{$^*$Work done during an internship at Google.} \quad Scott Wisdom$^{1}$  \quad John R.\ Hershey$^{1}$ \quad Aren Jansen$^{1}$ \quad Daniel P.\ W.\ Ellis$^{1}$}
\address{$^1$Google Research\\ $^2$Department of Computer Science, University of Illinois at Urbana-Champaign}
\begin{document}
\ninept
\maketitle
\begin{abstract}
Deep learning approaches have recently achieved impressive performance on both audio source separation and sound classification. Most audio source separation approaches focus only on separating sources belonging to a restricted domain of source classes, such as speech and music. However, recent work has demonstrated the possibility of ``universal sound separation'', which aims to separate acoustic sources from an open domain, regardless of their class. In this paper, we utilize the semantic information learned by sound classifier networks trained on a vast amount of diverse sounds to improve universal sound separation. In particular, we show that semantic embeddings extracted from a sound classifier can be used to condition a separation network, providing it with useful additional information. This approach is especially useful in an iterative setup, where source estimates from an initial separation stage and their corresponding classifier-derived embeddings are fed to a second separation network. By performing a thorough hyperparameter search consisting of over a thousand experiments, we find that classifier embeddings from clean sources provide nearly one dB of SNR gain, and our best iterative models achieve a significant fraction of this oracle performance, establishing a new state-of-the-art for universal sound separation.
\end{abstract}
\section{Introduction}
\label{sec:Intro}

A fundamental problem in machine hearing is deconstructing an acoustic mixture into its constituent sounds. This has been done quite successfully for certain classes of sounds, such as separating speech from nonspeech interference \cite{huang2014deep,wisdom2018differentiable} or speech from speech \cite{hershey2016deepclustering,Isik2016Interspeech09,Yu2017PIT,wang2018voicefilterTargetedVoiceSeparation,luo2019convTasNet}. However, the more general problem of separating arbitrary classes of sound, known as ``universal sound separation'', has only recently been addressed \cite{kavalerov2019universal}. An automated universal sound separation system has many useful applications, including better selectivity for assistive hearing devices, improved sound event classification for acoustic mixtures, and dynamic editing of sound and video recordings.

One major challenge for universal sound separation is the vast number of sound classes that may be encountered in the world. For any acoustic mixture, there is uncertainty about what types of sources are present. In this paper, we investigate how much universal sound separation can be improved if this uncertainty is reduced. In particular, we apply recent advances in sound classification \cite{LargeScaleAudioClassificationCNNs}, using the predictions of a sound classifier as embeddings that condition universal sound separation models. By doing so, we attempt to reduce the uncertainty about what types of sounds are contained in a mixture by providing additional information, and thus improve separation performance.

Capturing semantic information of high-dimensional data in a compact vector form (embedding) is essential for understanding and analyzing their underlying structure \cite{rudolph2016exponentialembeddings}. Conditional embeddings have been extensively used in various generation tasks for guiding a network to produce a semantically meaningful image given a context vector \cite{van2016CEmbImageGenerationPixelCNNdecoders, plummer2018CembImageText}, or just a class label \cite{mirza2014conditional}. Although recent works have been successful in finding semantically rich sound representations under an unsupervised setting \cite{UnsupervisedLearningSemanticAudioRepresentations, aytar2016soundnetsoundReprfromUnlabVideo}, there are few works that employ conditional embeddings towards other learning tasks on the audio domain. In \cite{kenter2019chiveTTSprosodic}, a conditional variational autoencoder is proposed in order to generate prosodic features for speech synthesis by sampling prosodic embeddings from the bottleneck representation. Specifically for separation tasks, speaker-discriminative embeddings are produced for targeted voice separation in \cite{wang2018voicefilterTargetedVoiceSeparation} and for diarization in \cite{cyrta2017SpeakerDiarSpeakerEmbeddings} yielding a significant improvement over the unconditional separation framework. Recent works \cite{seetharaman2019classconditionalembeddingsmusic, kumar2018musicActivityDetectionandSeparation} have utilized conditional embeddings for each music class in order to boost the performance of a deep attractor-network \cite{attractor} for music separation.

Despite the progress on exploiting conditional embeddings for audio separation tasks, there are still limitations. For example, embedding representations of the sources often need to be learned explicitly by auxiliary networks trained to predict the identity of the speaker or the sound classes contained in a mixture. In many real-world scenarios, the ground-truth class label of each source in a mixture is unknown or hard to annotate. To the best of our knowledge, this is the first work that proposes an end-to-end approach for conditioning universal sound separation without using any ground-truth class labels of the clean sources during training. Our specific contributions are as follows:
\begin{itemize}
    \item We use semantically rich audio representations produced by a sound classifier without assuming that we have access to the class labels of each source during training. 
    \item We propose several ways of conditioning a source separation network using the resulting semantic audio representations.
    \item We report state-of-the-art performance for universal sound separation by integrating conditional embeddings in an iterative separation framework. 
\end{itemize}

\section{Semantic Audio Representations}
\label{sec:SemanticAudiorepresentations}

\begin{figure*}[h]
    \centering
  \begin{subfigure}[h]{0.24\linewidth}
      \includegraphics[width=\linewidth]{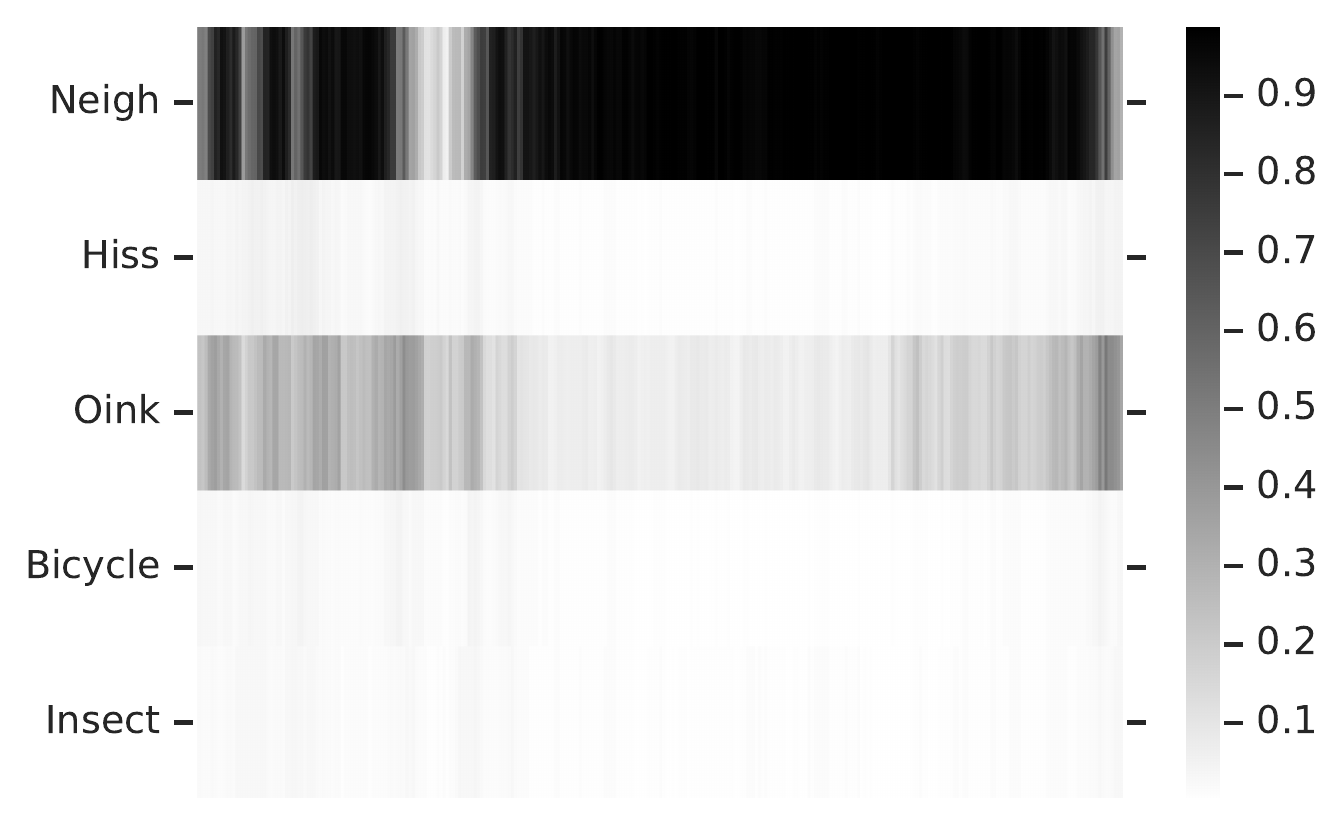}
      \caption{Source 1}
      \label{fig:source_embedding_1}
  \end{subfigure} 
  \begin{subfigure}[h]{0.24\linewidth}
      \includegraphics[width=\linewidth]{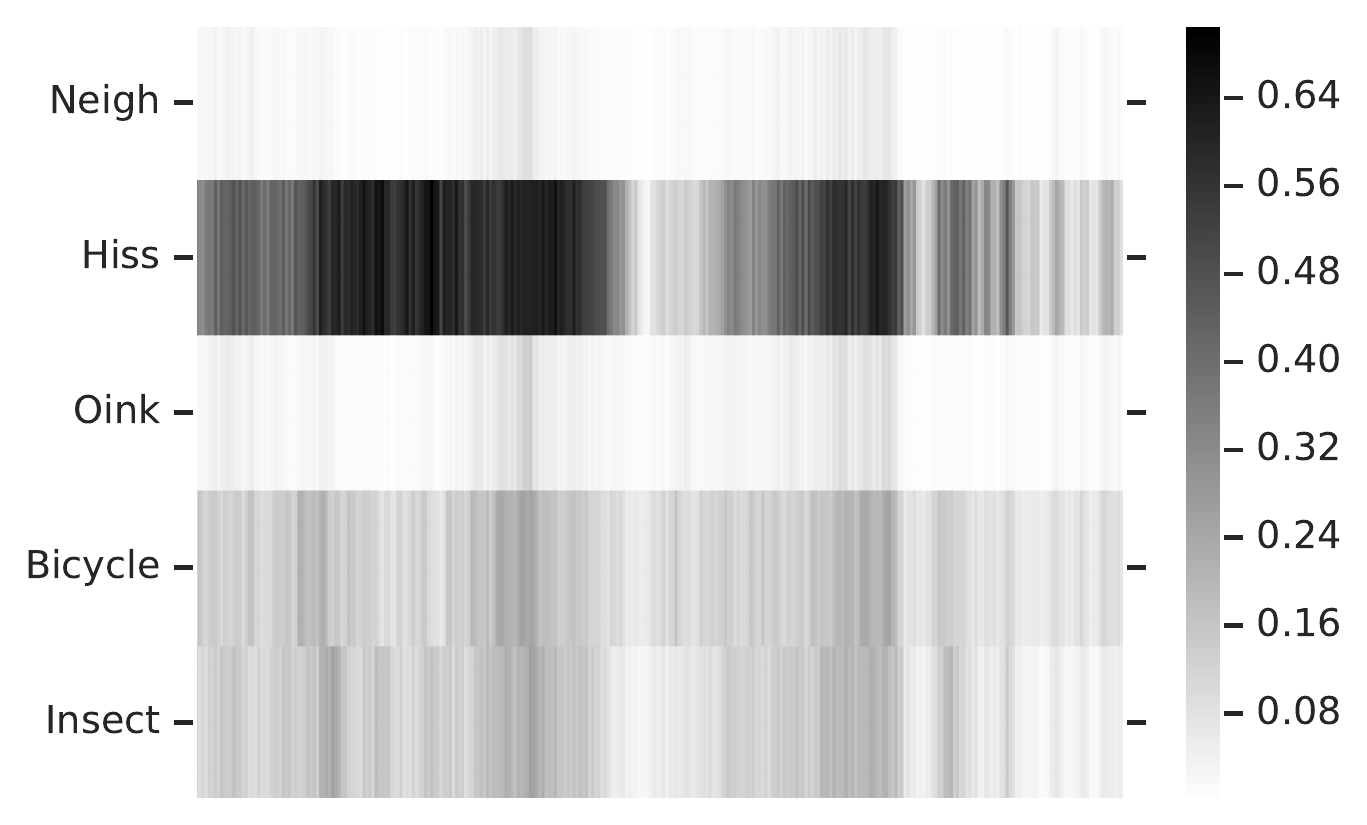}
      \caption{Source 2}
      \label{fig:source_embedding_2}
  \end{subfigure} 
  \begin{subfigure}[h]{0.24\linewidth}
      \includegraphics[width=\linewidth]{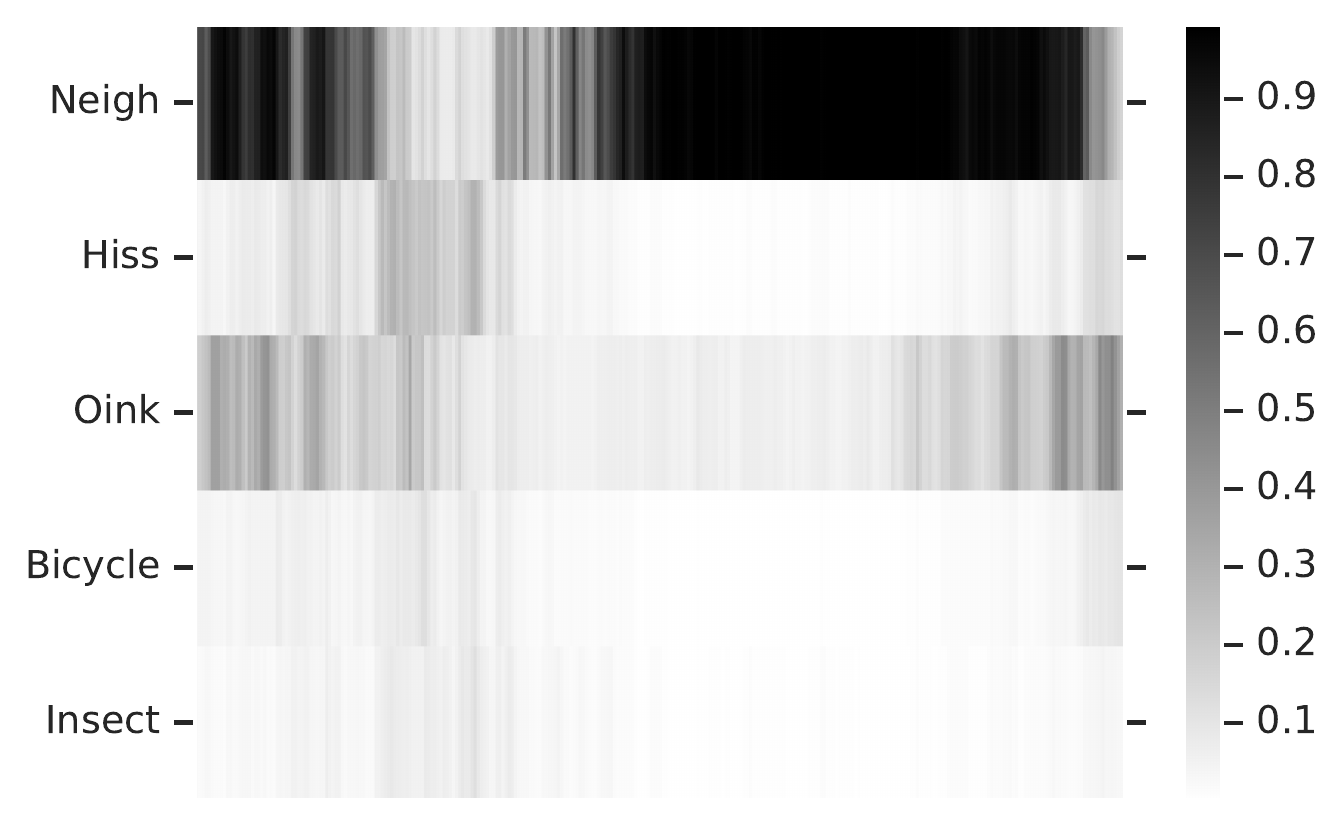}
      \caption{Mixture}
      \label{fig:mixture_embedding}
  \end{subfigure} 
  \begin{subfigure}[h]{0.24\linewidth}
      \includegraphics[width=\linewidth]{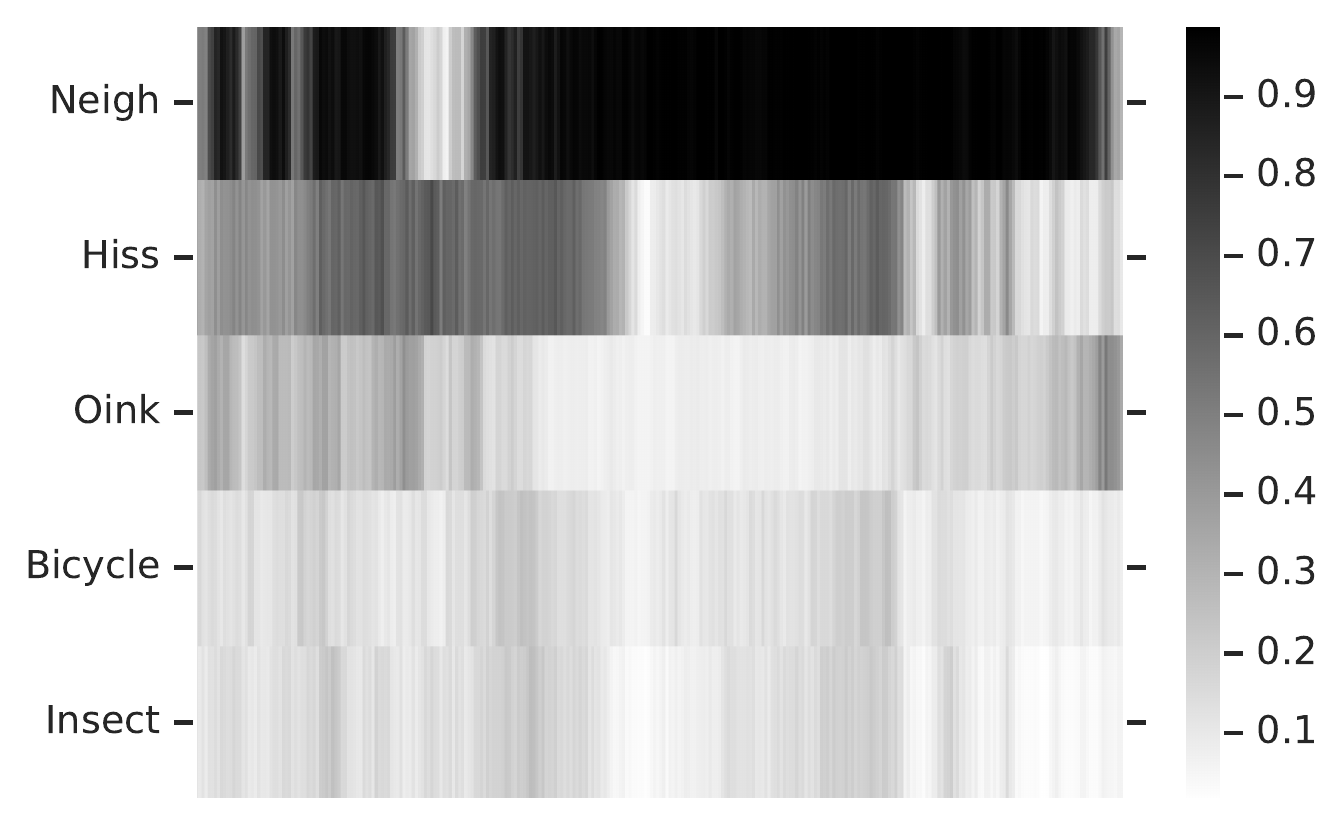}
      \caption{Soft-OR}
      \label{fig:softOR_embedding}
  \end{subfigure}
    \caption{Top 5 predicted classes over time for a mixture containing horse (a) and snake (b) sounds. Note that the corresponding probability tensor extracted from their mixture (c) does not enclose the semantic information of both sources, in contrast to the resulting soft-OR probability representation (d).}
    \label{fig:embeddings}
    \vspace{-15pt}
\end{figure*} 
\subsection{Extracting Embeddings from Sounds}
\label{sec:SemanticAudiorepresentations:Classifier}
In order to extract semantically meaningful representations of audio clips, we start with a pretrained sound classification network \cite{LargeScaleAudioClassificationCNNs, UnsupervisedLearningSemanticAudioRepresentations}. This model is trained using AudioSet \cite{AudioSet} to classify clips with multiclass labels drawn from 527 total sound classes.
In addition to the procedures described in \cite{LargeScaleAudioClassificationCNNs}, the sound classifier was trained using automatic temporal refinement of training labels, ontology ancestor smearing, and energy threshold silence labels. Also, unlike \cite{LargeScaleAudioClassificationCNNs}, we opt for a MobileNet-style architecture \cite{howard2017mobilenets} for its desirable computational footprint. This network consists of 10 layers of depthwise-separable convolutions, and we found this architecture to perform nearly as well as the larger architectures evaluated in \cite{LargeScaleAudioClassificationCNNs}. The MobileNet is fed patches of a 64-channel Mel-scale spectrogram with 25 ms window and 10 ms step. These shape $64\times96$ patches are extracted by a 96-frame sliding window with a hop of 1 frame (i.e.\ 10 ms).

The resulting conditional embedding that corresponds to an input signal $\textbf{s}$ is the predicted logits $\textbf{V} \in \mathbb{R}^{F \times 527}$, where $F$ is the number of predictions extracted by the sliding window. These logits are extracted from the last layer of the sound classifier, before application of the sigmoid activation. The conditional probability tensor of each sound class $c\in \{1, \cdots, 527\}$ contained in audio clip ${\bf s}$ over all frames $F$ is given by $P\left(c|\textbf{s}\right) = \sigma \left( \textbf{V}[:, c] \right) \in \left[0,1\right]^{F}$.

\subsection{Types of Embeddings for Source Separation}
\label{sec:SemanticAudiorepresentations:embedding_types}

Assuming that we know that each mixture contains $N$ sources, and considering that the clean sources $\textbf{s}_{1} \cdots \textbf{s}_{N}$ and the mixture $\textbf{m}$ have the same number of samples, we define the following types of semantic representations that we use for guiding our source separation networks: 
\begin{itemize}
    \item \textit{Mixture embedding}: the logits of the sound classifier after applying it to the mixture signal $\textbf{m}$: $\textbf{V}_{\textbf{m}} \in \mathbb{R}^{F \times 527}$.
    \item \textit{All embedding}: embeddings of mixtures and clean sources are concatenated.  That is, $\textbf{V}_{\textbf{s+m}} = \left[\textbf{V}_{\textbf{m}}, \textbf{V}_{\textbf{s}}\right] \in \mathbb{R}^{(N+1) \cdot F \times 527}$, where the sources embedding is the concatenation of the logits of the sound classifier after applying it to all clean source signals $\textbf{s}_{1} \cdots \textbf{s}_{N}$ individually: $\textbf{V}_{\textbf{s}} \in \mathbb{R}^{N \cdot F \times 527}$.
    \item \textit{Soft-OR embedding}: first we compute the soft-OR operation on the probability space from all available sources, and then we compute the inverse sigmoid function $\sigma^{-1}(\cdot)$ of the result, to go back to the logits space. For a single class $c$, we define the soft-OR operation over $N$ sources as:
    \begin{equation}
        P_{or}\left(c|\textbf{s}\right) = 1 - \prod_{i=1}^{N} (1 - P\left(c|\textbf{s}_i\right)).
    \end{equation}
    Intuitively, the soft-OR is the probability that at least one of $N$ biased coins comes up heads.
    To convert back to logits, we define the soft-OR embedding as $\textbf{V}_{or}[:, c] = \text{logit} \left(P_{or}\left(c|\textbf{s}\right)\right)$, where $\text{logit}(P)=\text{log}(P / (1-P))$.
\end{itemize}
Note that the ``soft-OR'' and ``all'' embeddings require access to the sources (either clean references or estimated) in order to be extracted. 
Examples of these different semantic representations on the probability space of two sources as they are predicted by the sound classifier 
are shown in Figure \ref{fig:embeddings}.
Ideally the semantic representation of the mixture should resemble the soft-OR embedding ( $\textbf{V}_{\textbf{m}} \approx \textbf{V}_{or}$), because if a sound exists in one or more of the sources, then it must exist in the mixture, and the embedding would then represent the presence or absence of the sounds to be separated.

\section{Audio Source Separation Frameworks}
\label{sec:Models}

\subsection{Time-Domain Source Separation}
\label{sec:Models:Timedomain}
We adopt the  TDCN++ model and its iterative version \mbox{iTDCN++}, for universal source separation, based on their performance in \cite{kavalerov2019universal}. The TDCN++ model consists of analysis and synthesis basis transforms which encode and decode the signal, respectively, as well as a masking-based separation module that consists of stacked blocks of dilated convolutions and dense layers similar to \cite{luo2019convTasNet}. In the iterative version (iTDCN++), the process of estimating the source signals is repeated twice. The mixture audio $\textbf{m}$ and output estimates $\hat{\textbf{s}}^{(1)}$ of the first TDCN++ module serve as input for the second TDCN++ module, which produces the final separation estimates $\hat{\textbf{s}}^{(2)}$. Both separation networks are trained using permutation-invariant \cite{Yu2017PIT} negative signal to noise ratio (SNR), given by 
\begin{equation}
\mathcal{L}_{sep} = -SNR \left( \mathbf{s}_{p^*}, \hat{\textbf{s}}\right) = -10 \log_{10}
    \left(
    \| \textbf{s}_{p^*}\|^2
    /
    \| \textbf{s}_{p^*} - \hat{\textbf{s}}\|^2
    \right)
\end{equation}
where $p^*$ denotes the best possible permutation of the sources $\textbf{s}$ given some model predictions $\hat{\textbf{s}}$. For the iTDCN++ model, we use the separation losses for both iterations, namely, $\mathcal{L}_{isep} = \mathcal{L}^{(1)}_{sep} + \mathcal{L}^{(2)}_{sep}$.

\begin{figure}[!htb]
    \centering
      \includegraphics[width=\linewidth]{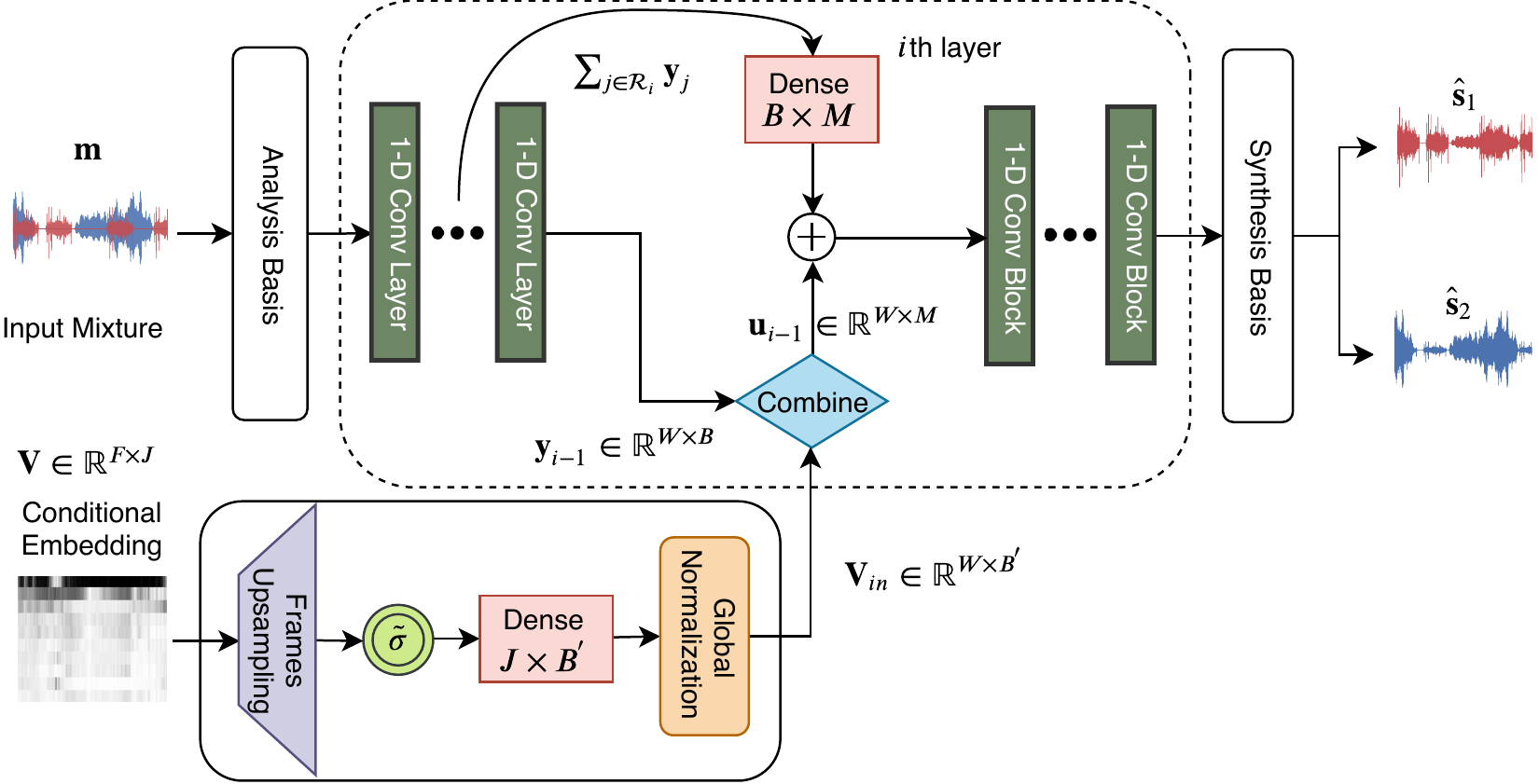}
      \caption{Integrating conditional embedding information to the $i$th layer of a TDCN++ separation module.}
      \label{fig:IntegratingConditionalEmbeddings}
\end{figure} 

\begin{figure*}[!h]
    \centering
  \begin{subfigure}{0.31\linewidth}
  \flushleft
      \includegraphics[width=\linewidth]{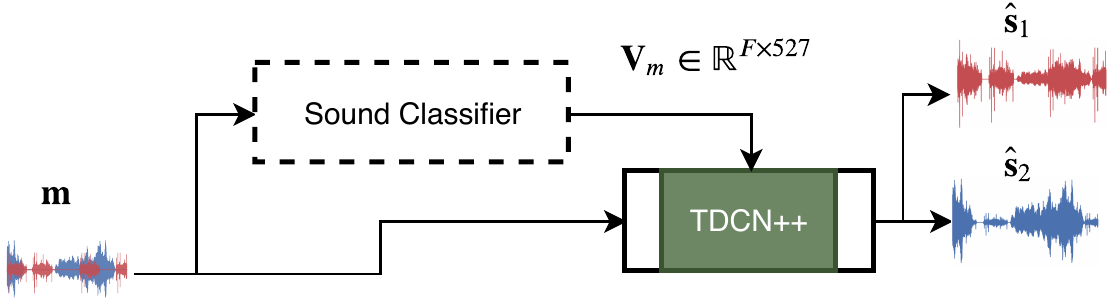}
      \caption{With pretrained mixture embedding}
      \label{fig:baseline_mixture}
  \end{subfigure} 
  \begin{subfigure}{0.31\linewidth}
  \flushright
      \includegraphics[width=0.81\linewidth]{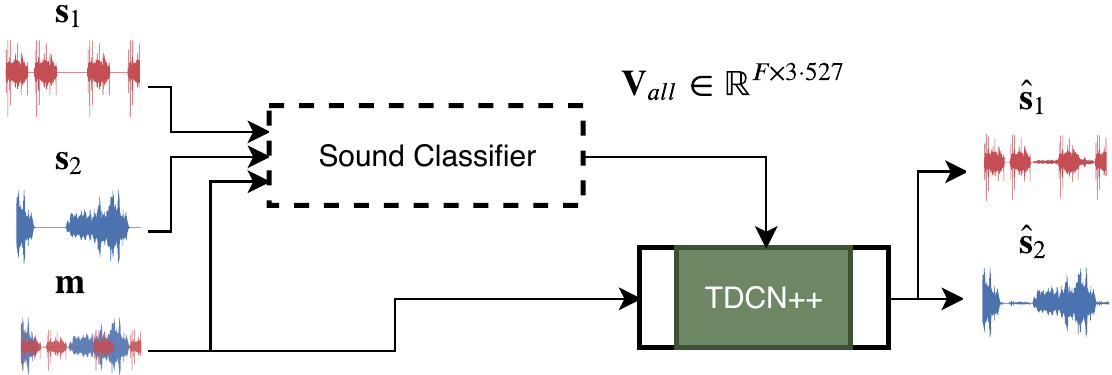}
      \caption{With pretrained oracle embeddings.}
      \label{fig:oracle_embeddings}
  \end{subfigure} 
  \begin{subfigure}{0.37\linewidth}
    \flushright
      \includegraphics[width=0.9\linewidth]{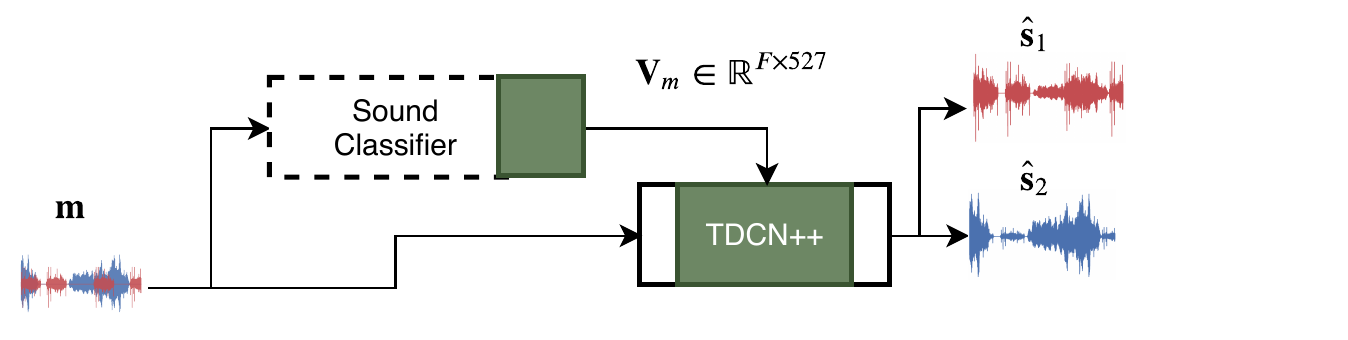}
      \caption{With fine-tuned embeddings.}
      \label{fig:e2e}
  \end{subfigure} \\
  \begin{subfigure}{\linewidth}
  \centering
      \includegraphics[width=0.7\linewidth]{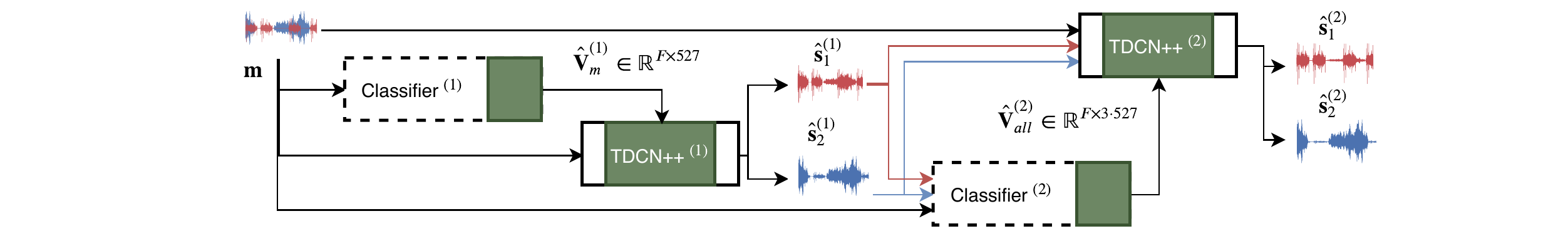}
      \caption{Iterative with fine-tuned embeddings.}
      \label{fig:iterative}
  \end{subfigure}
    \vspace{-10pt}
    \caption{High-level block diagrams of the various separation network settings that we consider. Dark green blocks indicate trainable weights.}
    \label{fig:high_level_experiments}
    \vspace{-10pt}
\end{figure*} 
\vspace{-15pt}
\subsection{Integrating Conditional Embeddings}
\label{sec:Models:integrating}
For the $i$th layer of a TDCN++ separation network, the input from the previous layer is $\textbf{y}_{i-1} \in \mathbb{R}^{W \times B}$, where $W$ is the number of frames and $B$ is the number of channels. Residual connections from previous layers are added together $\sum_{j \in \mathcal{R}_i} \textbf{y}_{j}$, where $\mathcal{R}_i$ is the set of previous layer indices. This result is input to a $1 \times 1$ convolutional layer that converts the channel dimension to $M$. To integrate a conditional embedding $\textbf{V} \in \mathbb{R}^{F \times J}$, the following steps are taken (see Figure \ref{fig:IntegratingConditionalEmbeddings}):
\begin{enumerate}
    \item The embedding is resampled in order to match the time dimensions of the current layer's activations $W$ by repeating frames.
    \item 
    We apply either a) a regular sigmoid function $\sigma \left(x \right) = \nicefrac{1}{(1 + e^{-x})}$ or b) a trainable sigmoid $\sigma \left(x \right) = \nicefrac{\alpha}{(1 + e^{\beta \left( x_0 - x \right)})}$. 
    with trainable parameters $\alpha$, $\beta$, $x_0$ \cite{nelder1961fittinglogisticcurve}.
    \item The channels dimensions of the input conditional embedding $J$ are matched to $B'$ through a $1 \times 1$ convolutional layer and the whole tensor is fed to a global normalization layer in order to match the statistics of the activations.
    \item The resulting input conditioning tensor $\mathbf{V}_{in} \in \mathbb{R}^{W \times B'}$ is combined with the activations from the $i-1$th layer either a) by concatenation $u_{i-1} = \left[ \textbf{V}_{in}, \enskip \textbf{y}_{i-1}  \right]$ or b) by gating $u_{i-1} = \textbf{V}_{in} \odot \textbf{y}_{i-1}$ (the we enforce that $B = B'$). The number of channels from the output of the dense layer of the residual connections is $M = \left(B + B' \right)$ and $M = B$ for the two cases, correspondingly.
\end{enumerate}
This process enables the integration of semantic information in any layer of the separation module. In this work, we consider two cases where the semantic information is either integrated in only the first layer of TDCN++ module, or in all layers.
\subsection{Pretrained Embeddings}
\label{sec:Models:pretrained} 
For this setting, our goal is to determine if embeddings from the pretrained sound classifier can improve separation performance. We use the pretrained sound classifier described in Section \ref{sec:SemanticAudiorepresentations:Classifier} and we freeze its parameters. For each input mixture, the sound classifier predicts a corresponding embedding tensor $\textbf{V}$ which is integrated into the TDCN++ model as explained in the previous section. We consider two variations of this experiment. In the first case, which is also depicted in Figure \ref{fig:baseline_mixture}, we suppose that we can only extract the mixture embedding $\textbf{V}_{\mathbf{m}}$ during training and inference times. In the other case, we consider an oracle scenario in which the network has access to the clean sources as shown in Figure \ref{fig:oracle_embeddings}. Thus, the sound classifier produces embeddings for both the sources and the mixture. The concatenated tensor $\textbf{V}_{\mathbf{all}}$ is now the conditional embedding which guides the TDCN++ separation network. In both baseline and oracle cases, the training loss remains the same as the conventional time-domain source separation setup $\mathcal{L}_{sep}$. 
\subsection{Fine-tuned Embeddings}
\label{sec:Models:e2e} 
As discussed in Section \ref{sec:SemanticAudiorepresentations:embedding_types}, the pretrained embedding corresponding to the mixture signal $\textbf{V}_{\mathbf{m}}$ might not contain significant information for all the existing sources. To avoid this problem, we propose a fine-tuning scheme as shown in Figure \ref{fig:e2e}, where the last $1$ or $3$ layers of the sound classifier are trainable. This provides the sound classifier with more degrees of freedom which can be used to refine the conditional mixture embedding $\hat{\textbf{V}}_{\mathbf{m}}$ in order to make it more informative for
the separation network. The whole architecture is trained in an end-to-end fashion using only the separation loss $\mathcal{L}_{sep}$.
\subsection{Fine-tuned Embeddings for the Iterative Model}
\label{sec:Models:iterative} 
Although the end-to-end fine-tuning can be used to refine the mixture embedding, conditional embeddings obtained from estimates of the clean sources could contain more precise information. For the iterative estimation of the conditional embeddings, the iTDCN++ is used as the separation network architecture as it is shown in Figure \ref{fig:iterative}. This extends the aforementioned end-to-end approach in order to predict estimates of the clean sources $\hat{\textbf{s}}^{(1)}$. These estimates, as well as the mixture $\textbf{m}$, are fed to a second sound classifier 
to extract embeddings for the source estimates and the mixture $\hat{\textbf{V}}_{\mathbf{all}}^{\left(2\right)}$. These embeddings condition a second source separation subnetwork TDCN++$^{(2)}$ which produces the final source estimates $\hat{\textbf{s}}^{(2)}$. The whole architecture is optimized with $\mathcal{L}_{isep}$ similar to an unconditioned iTDCN++ as described in Section \ref{sec:Models:Timedomain}.   

\begin{table*}[ht]
    \centering
    \begin{tabular}{l|l|cc|cc|cc}
    
    \hlinewd{1pt}
    \toprule
    & & \multicolumn{2}{c|}{Embeddings} & \multicolumn{2}{c}{STFT} & \multicolumn{2}{c}{Learned}  \\
    & Method & Type & Fine-tuning & Val. & Test & Val. & Test \\
    \hlinewd{1pt}
    \multirow{2}{*}{Baselines }
        & TDCN++ with no embeddings \cite{kavalerov2019universal} & - & -& 
        9.9 & 9.1 & 9.1 & 8.5
        \\
        & iTDCN++ with no embeddings \cite{kavalerov2019universal} & - & - &
        10.6 & 9.8 & 9.3 & 8.7
        \\
    \hline
    \hlinewd{1pt}
    \multirow{6}{*}{Proposed}
        & Pretrained embeddings \& TDCN++ & mixture & - &
        10.3 & 9.4 & 9.4 & 8.6
        \\
        & Fine-tuned embeddings \& TDCN++ & mixture & \ding{51} &
        10.2 & 9.4 & 9.3 & 8.5
        \\
        & Guided fine-tuned embeddings \& TDCN++ & mixture & \ding{51} &
        10.3 & 9.4 & 9.4 & 8.6
        \\
        \cline{2-8}
        & Pretrained embeddings \& iTDCN++ & all & - &
        10.8 & 9.9 & 9.9 & 9.0
        \\
        & Fine-tuned embeddings \& iTDCN++ & all & \ding{51} &
        {\bf 11.1} & 10.1 & {\bf 10.1} & {\bf 9.2}
        \\
        & Guided fine-tuned embeddings \& iTDCN++ & all & \ding{51} &
        {\bf 11.1} & {\bf 10.2} & 10.0 & 9.1
        \\
    \hlinewd{1pt}
    \multirow{3}{*}{Oracles} & \multirow{2}{*}{Pretrained embeddings \& TDCN++}
        & all & - &
        11.3 & 10.6 & 11.0 & 10.2
        \\
        & & soft-OR & - &
        11.4 & 10.6 & 10.7 & 10.1
        \\
    \cline{2-8}
        & STFT binary mask & - & - &
        16.8 & 16.2 & - & -
        \\
    
    \bottomrule
    \end{tabular}
    \caption{Best model performance in terms of SI-SDRi (dB).}
    \label{tab:best_models}
    \vspace{-9pt}
\end{table*}

\subsection{Guided Fine-tuned Embeddings for the Iterative Model}
\label{sec:Models:Guidediterative} 
In the previous conditions, the last few layers of the classifiers are fine-tuned based on their contribution to the final separation loss. Building upon that, we use the oracle embeddings extracted from the clean sources of the training set as targets in order to guide the classifiers towards producing similar representations. In order to do so we use a sigmoid cross-entropy loss
\begin{equation}
CE\left(\textbf{v}_1, \textbf{v}_2  \right) = - \mathbb{E}_{\sigma \left(\textbf{v}_1\right)}\left[\operatorname{log}_2\sigma(\textbf{v}_2)\right]
\end{equation}
at the output of each classifier and add them to the total loss. Specifically, we would like to ensure that the mixture embeddings $\hat{\textbf{V}}_{\mathbf{m}}^{\left(1\right)}$ and $\hat{\textbf{V}}_{\mathbf{m}}^{\left(2\right)}$ stay close to the soft-OR embedding $\textbf{V}_{or}$ and that the predicted source embeddings $\hat{\textbf{V}}_{\mathbf{s}}^{\left(2\right)}$ stay close to the oracle embeddings $\textbf{V}_{\mathbf{s}}$. Thus, the total loss for this framework is: 
\[\mathcal{L}_{isep} + CE \left( \mathbf{V}_{or}^{p^*}, \hat{\textbf{V}}^{(1)}_m\right) + CE \left( \mathbf{V}_{or}^{p^*}, \hat{\textbf{V}}^{(2)}_m\right) + CE \left( \mathbf{V}_{s}^{p^*}, \hat{\textbf{V}}^{(2)}_{s}\right)\]

\section{Experiments}
\label{sec:Experiments}

\vspace{-7.5pt}
\subsection{Datasets}
For our universal sound separation experiments we use the dataset introduced in \cite{kavalerov2019universal}. This dataset contains a wide variety of sound classes and is sources from the Pro Sound Effects Library database \cite{prosound}. Each input source consists of a single-channel three-second clip sampled at $16$kHz, whereas the generated mixtures contain random combinations of all the available sources. Overall, there are $14,049$ training mixtures ($11.7$ hours), $3,898$ validation mixtures ($3.2$ hours), and $2,074$ test mixtures ($1.7$ hours). We focus on the two-source task in this paper, though our methods could be easily extended to more sources.

\vspace{-7pt}
\subsection{Training and Evaluation Details}
All models are implemented in TensorFlow \cite{tf_short}. Models are trained using the Adam \cite{adam} optimizer on a single NVIDIA Tesla V100 GPU for around $3$ million steps with a learning rate of $10^{-4}$ and batch size of $2$. The window and hop sizes are $5$ ms and $2.5$ ms, respectively, in both the analysis and synthesis layers, since these are the optimal parameters reported by \cite{kavalerov2019universal}. The learnable basis uses $256$ coefficients, while the STFT basis uses $65$. A ReLU is applied to learnable basis coefficients, while a magnitude operation is used for STFT. For the integration of the conditional embeddings as described in Section \ref{sec:Models:integrating}, we set the number of channels in the intermediate layers of TDCN++ to $B=128$ and for the conditional embeddings to $B'=128$. Each sound classifier outputs an embedding with $F=301$ time frames and $J=527$ classes for each $3$ second input.   

The performance of all models is evaluated using scale-invariant signal-to-distortion ratio improvement (SI-SDRi) \cite{Isik2016Interspeech09, leroux2018sdr, luo2019convTasNet}. Formally, SI-SDR between an estimated signal $\hat{\textbf{s}}$ and the ground truth signal $\textbf{s}$ can be expressed as:
\begin{align}
    \text{SI-SDR}(\textbf{s}, \hat{\textbf{s}}) =  10 \log_{10} \left( \| \gamma \textbf{s}\|^2 / \| \gamma \textbf{s} - \hat{\textbf{s}}\|^2 \right)
\end{align}
where $ \gamma = \argmin_{a} \| a \textbf{s} - \hat{\textbf{s}}\|^2 = \langle \textbf{s}, \hat{\textbf{s}}\rangle /\|\textbf{s}\|^2$, and $\langle \textbf{a},\textbf{b} \rangle$ denotes the inner product.  SI-SDRi is the difference between the SI-SDR of the estimated signal and that of the input mixture signal.

\vspace{-7pt}
\subsection{Results}
\label{sec:ExperimentsResults}

Table \ref{tab:best_models} shows the results of our best models for each condition we considered. Notice that iterative models that use embeddings perform the best in terms of the proposed methods. Consistent with \cite{kavalerov2019universal}, the STFT basis outperforms the learnable basis.
Also note that the performance of the TDCN++ with pretrained mixture embedding is quite similar to TDCN++ with the fine-tuned counterpart. In contrast, the iterative iTDCN++ improves when fine-tuned.
This suggests that semantic representations extracted from estimates of the clean sources are able to better condition the separation network.
Generally, the guided models that use a cross-entropy loss on predicted embeddings do not affect separation performance significantly, but at least the overall model can be trained to perform better classification without affecting separation.
The oracle methods at the bottom of the table indicate the total possible gain from conditioning on sound classifier embeddings. However, even these oracle embedding methods are below the oracle STFT binary mask, indicating that more improvement is still possible.

In order to find the best configuration parameters for the networks, we perform a large-scale empirical study where we trained over $1000$ models. Given the large scale of this study, we share some of the most prominent experimental findings when sweeping over different configurations of the integration parameters as described in Section \ref{sec:Models:integrating}. 
Time-varying embedding predictions, repeated out to match the number of basis frames, outperformed a time-invariant embedding computed by taking the mean across frames in probability space.
Fixed and trainable sigmoid functions scored higher than conditioning on the raw logits. 
For the number of channels $B' \in \{65, 128, 256\}$, $128$ generally yielded the best results. 
Between the concatenation and gating integration methods as described in section \ref{sec:Models:integrating}, concatenation surpassed gating most of the time.
Despite our initial intuition that by feeding $\textbf{V}_{in}$ to all convolutional blocks of the TDCN++ would better guide the network towards separation, it instead led to overfitting. Thus, combining $\textbf{V}_{in}$ only with the analysis basis coefficients consistently led to better performance.

\section{Conclusion}
\label{sec:Conclusion}

In this paper we have shown that universal sound separation performance can be improved by conditioning on the predictions of sound classifiers trained on large amounts of diverse audio. Our best models use an iterative setup, where sources are initially separated and classified, then fed to a second separation stage. These models achieve a new state-of-the-art benchmark for universal sound separation, while also providing useful classification outputs.

In future work, we will investigate using other types of conditioning information. Classification predictions are a coarse summary of the information in a signal, and we expect that combining them with richer signals such as video could further improve performance. Also, we intend to try this task with larger amounts of data and more human annotation, such as ground-truth class labels.

\vfill
\pagebreak 
\bibliographystyle{IEEEtran_nourl}
\balance
\bibliography{refs}

\end{document}